\documentclass[namedreferences]{solarphysics}
\usepackage[optionalrh]{spr-sola-addons} 
\usepackage{graphicx}        
\usepackage{color}           
\usepackage{url}             

\begin{document}
\begin{article}
\begin{opening}

\title{The multifrequency Siberian Radioheliograph}

\author{S.V.~Lesovoi,~A.T.~Altyntsev,~E.F.~Ivanov,~A.V.~Gubin}

\runningauthor{S.V.~Lesovoi~et~al.}

\runningtitle{The Siberian Radioheliograph}

\institute{Institute~of~Solar-Terrestrial~Physics,
126a~Lermontov~Str., Irkutsk, 664033, P.O.~Box~4026,Russia}

\date{April 16, 2012}

\begin{abstract}
The 10-antenna prototype of the multifrequency Siberian radioheliograph is described. 
The prototype consists of four parts: antennas with broadband front-ends, 
analog back-ends, digital receivers and a correlator. The prototype  
antennas are mounted on the outermost stations of the \textit{Siberian Solar Radio 
Telescope} (SSRT) array. A signal from each antenna is transmitted 
to a workroom by an analog fiber optical link, laid in an underground tunnel. After mixing, 
all signals are digitized and processed by digital receivers before
the data are transmitted to the correlator. The digital receivers and the correlator are 
accessible by the LAN. The frequency range of the prototype is from 4 to 8~GHz. Currently the frequency switching observing mode is used. The prototype data include both circular polarizations at a number of frequencies given by a list. This prototype is the first stage of the multifrequency Siberian radioheliograph development. It is assumed that the radioheliograph will consist of 96 antennas and will occupy stations of the West-East-South subarray of the SSRT. The radioheliograph will be fully constructed 
in autumn of 2012. We plan to reach the brightness temperature sensitivity about 100~K for the snapshot image, a spatial 
resolution up to 13~arcseconds at 8~GHz and polarization 
measurement accuracy about a few percent.

First results with the 10-antenna prototype are presented of observations of solar microwave bursts. The prototype abilities to estimate source size and locations at different frequencies are discussed. 

\end{abstract}
\end{opening}

\section{Introduction}

The approach to design a new kind of solar radiotelescope is well known for many years \cite{GaryFASR}. Nevertheless, there are still no solar radio telescopes able to obtain data matching with the excellent data of satellites. But it is very likely that such problems in solar physics as particle acceleration, coronal heating and coronal magnetic field measurement cannot be solved without studying the radio emission of the Sun. There are a few projects working toward the next generation of solar radio telescopes. The \textit{Frequency Agile Solar Radiotelescope} (FASR) subsystem testbed \cite{LiuPacific} is operating as a prototype of the FASR system. The first FASR prototype observation of the zebra pattern radio burst \cite{ChenApJ} is a very promising result. It demonstrates a fruitful approach of simultaneous spectral and interferometric observations. The \textit{Chinese Spectral Radioheliograph} \cite{Yan2011} perhaps is the only new instrument under construction today. But both above mentioned instruments are rather in the decimetric frequency range now. The prototype of the Siberian radioheliograph operates at microwaves. The main goal of the 10-antenna prototype is to study some technical questions on the development of a multifrequency radioheliograph and to estimate its cost. The results of the study will be used in the upgrade of the \textit{Siberian Solar Radio Telescope} (SSRT) to such a multifrequency radioheliograph. The SSRT is a solar dedicated radio telescope with imaging utilizing both frequency scanning and Earth rotation \cite{GrechnevSolPh}. The following design issues have been resolved during the 10-antenna prototype project: choice of a suitable dual polarization feed; method of broadband RF signal transmission from the antennas to a control room; way for delay tracking and fringe stopping; method of data transmission from digital receivers to the correlator and design of the correlator; and finally, choice of a suitable data format.
The prototype consists of four hardware parts and acquisition/control software. The main hardware parts are antenna equipped with a dual polarization broadband analog front-end with optical link, a downconverter, a digital receiver and the correlator. The software consists of the embedded applications running under $\mu$C/OS-II, the Linux application controlling the observation and some Python and PHP scripts.
The next milestone of the multifrequency radioheliograph project is first observations in autumn of 2012. The 96-antenna heliograph will be constructed at this time. It is assumed that only half of T-shape West-East-South 192-antenna array of the SSRT will be utilized. It would be useful to utilize all antenna stations of this T-shape array however it is impossible today due to the limited financial support of this project.

\section{Instrumentation}
\subsection{System design}
\begin{figure}
\caption{Block diagram of the 10-antenna prototype of the multifrequency Siberian radioheliograph. The labels above arrows show bandwidths, gain and sample rate.} \label{bd_srh}
\centerline{\includegraphics[width=\maxfloatwidth]{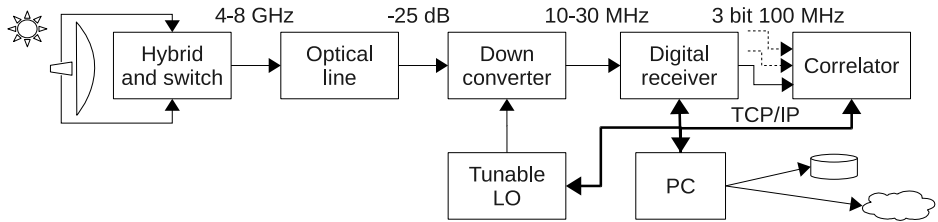}}
\end{figure}
Figure~\ref{bd_srh} shows the block diagram for the whole system design. The prototype utilizes ten of the outermost antenna stations of the SSRT: four antennas at the South subarray and three antennas at both West and East subarrays. The shortest baseline is 4.9~m long, the longest one is 622.3~m (in the East-West direction). The signals from antennas are transmitted to the workroom by the analog optical link. All cables are around 375 m long. The outdoor part of the cable has length of around 5~m while most of the cable is laid in the SSRT underground tunnel.The frequency range of the prototype is from 4~GHz up to 8~GHz. The snapshot image is obtained in the upper side band near the local oscillator frequency: $\nu_{LO}+\nu_0\pm\Delta\nu$, where $\nu_0 = 17$~MHz and $\Delta\nu = 5$~MHz. In fact, the digital receiver and the correlator are field programmable gate array~(FPGA) intellectual property~(IP) cores excluding the digital receiver's analog-to-digital converter~(ADC). The data are transmitted from digital receivers to the correlator by low voltage differential signaling~(LVDS) pairs.

The raw data of the prototype consist of 10 autcorrelations (fluxes) and 45 complex cross-correlations (visibilities). These data are stored on a hard disk, and single dish fluxes and visibilities averaged over some range of spatial frequencies are provided online\footnote[1]{http://badary.iszf.irk.ru/prototype\_10.php}.

\subsection{Broadband antenna}
\begin{figure}
\caption{Block diagram of the broadband antenna. The antenna is equipped with a dual linear polarization feed. Therefore the quadrature hybrid is used to obtain circular polarizations. The large insertion loss of the optical modulator is compensated by the broadband 30~dB amplifier just before it in the signal path.} \label{bd_antenna}
\centerline{\includegraphics[width=\maxfloatwidth]{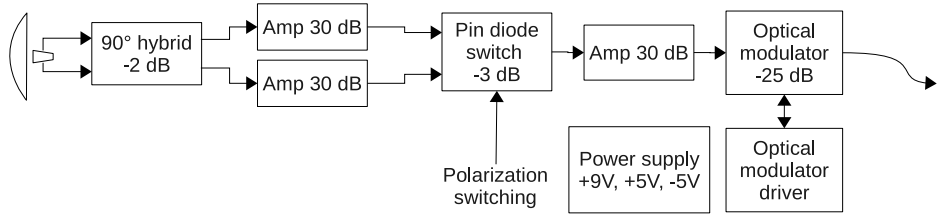}}
\end{figure}

The broadband antennas were newly developed for the 10-antenna prototype. This antenna consists of the metal parabolic dish and the dual linear polarization horn feed. The quadrature hybrid coupler is used to convert linear polarizations to circular ones. Although the beam patterns of such feed are slightly different for right and left circular polarizations, the difference is less than $1\%$ in the whole frequency range. The insertion loss of this feed is much less with respect to an equivalent sinuous feed and the horn feed is more reliable device that works in a wider temperature range. The frequency range of the feed is 4.5 to 9.0~GHz with VSWR (Voltage Standing Wave Ratio) is less than 2.0, the isolation between different polarizations is about 25~dB. The antenna gain is 35.1 (4.5~GHz) to 40.2 (9.0~GHz). The antenna front-end (shown in Figure~\ref{bd_antenna}) consists of a quadrature hybrid coupler, 30~dB preamplifiers, pin-diode switch, 30~dB amplifier and optical transmitter. The last amplifier is to compensate large insertion loss in the optical modulator. The typical loss is -25~dB to -30~dB due to extremely low RF to optical conversion efficiency. The spur free dynamic range of the optical modulator is about 45~dB for 4~GHz bandwidth. The gain of the analog back-end is about 35 dB and the gain flatness is about 3~dB. 

\subsection{Downconverter}
\begin{figure}
\caption{Block diagram of the downconverter. The labels show the optical wavelength, the RF bandwidth and the IF bandwidth.} \label{bd_downconverter}
\centerline{\includegraphics[width=\maxfloatwidth]{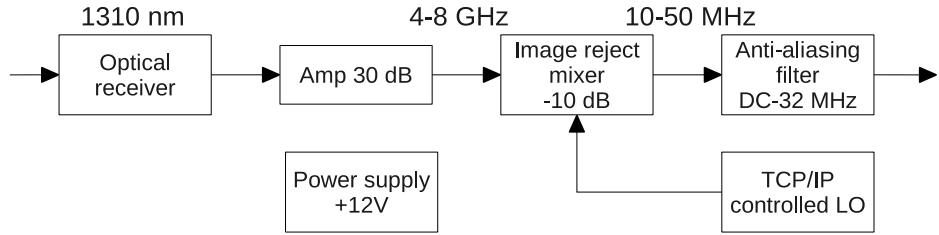}}
\end{figure}

The optical signal is converted to RF using an optical receiver (Figure~\ref{bd_downconverter}). The RF signal is amplified and downconverted to an IF range of 10 to 50~MHz using an image reject mixer~(IRM). The IRM consist of a I/Q mixer and a hybrid coupler. The upper side band is used. The frequency range of the IRM is 4.0 to 8.0~GHz. The image rejection is better than 25~dB. The LO to RF isolation is at least 40 dB and the LO to IF isolation is about 30 dB. The RF signal is filtered by amplifiers whose passband is 4-8~GHz, because no need to use the RF filter before the IRM. The frequency agile generator, controlled by TCP/IP data transmission protocol, is used as the local oscillator. Because the IRM input LO power is about $15$~dBm the output power of the LO has to be relatively high (23-25~dBm).

\subsection{Digital receiver}
\begin{figure}
\caption{Block diagram of the digital receiver. The FIR(t) and delay(t) are to compensate the geometrical delay. The NCO(t) is to fringe stopping. The truncating has to be done to prevent the digital receiver overflowing.} \label{bd_dreceiver}
\centerline{\includegraphics[width=\maxfloatwidth]{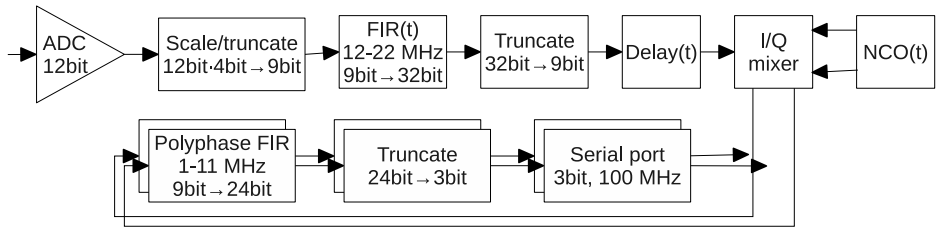}}
\end{figure}

The digital receiver utilizes the Altera DSP development kit Stratix II 2S60. The receiver is aimed to convert the analog IF signal to digital data and to process these data before  correlation. Firstly, a passband of the signal has to be formed. A digital 64-th order FIR filter is used to form it with a passband of 12 to 22 MHz. Because the sample rate of the ADC is of 100~MHz, while the desired step of the delay compensation is of 10 times smaller (1~ns) the FIR filter is also used to implement the fractional sample delay by a suitable set of impulse response functions of the filter is used to do it. Adjacent impulse response functions correspond to a time delay of 1~ns. Further, the signal is delayed by the delay line with step of one ADC clock (10 ns). In other words, one can define the geometrical delay $\tau^{k}_{g}=n\tau^{k}_{10}+m\tau^{k}_{1}+\delta\tau^{k}_{g}$, where $k$ is an antenna number, $\tau^{k}_{10}$ and $\tau^{k}_{1}$ are integer parts and $\delta\tau^{k}_g$ is fractional part of $\tau^{k}_g$ expressed in nanoseconds. In such a case $m$ is a number of the filter impulse response function, $n$ is a number of the delay line (shift register) tap.  The fractional part of the geometrical delay $\delta\tau^{k}_g$ leads to a phase error to be corrected \cite{Thompson}. The whole necessary phase correction is defined as $\omega_{LO}\tau^{k}_{i}+\omega_{RF0}\delta\tau^{k}_{g}$, where $\omega_{LO}$ is the local oscillator angular frequency, $\tau^{k}_{i} = n\tau^{k}_{10}+m\tau^{k}_{1}$ and $\omega_{RF0}$ is the central RF angular frequency. Both members of the phase correction are compensated by the numerically controlled oscillator (NCO) phase tuning (Figure~\ref{bd_dreceiver}). The frequency band of the digital receiver is 10~MHz. The delay error at the correlator input $\delta\tau^{k}_g - \delta\tau^{l}_g$ varies in the range $\pm1$~ns, where $k, l$ are antenna numbers. Thus the phase error within the frequency band is about 1$^\circ$ and the resulting correlation loss of about 1\% is tolerable. The NCO is used not only to correct the phase but also to produce IQ signals and downconvert IF signals. The IQ signals, to be transmitted to the correlator, are produced by the digital IQ mixer. The IF signal bandwidth on the output of the polyphase low-pass filter is in range 1-11~MHz. So it is possible to serialize the IQ data and transmit them to the correlator by serial lines. A sample rate of about 100 MHz is required to serialize a 3-bits signal with bandwidth about 10~MHz. Due to the relatively low sample rate, simple LVDS pairs are used.

\subsection{Correlator}
The correlator of the 10-antenna prototype is a FPGA IP core consisting of 55 complex correlators, and an interface to the embedded Nios~II processor. The number of complex visibilities for the 10-antenna array is $N(N-1)/2 =$~45 and 10 correlators are used as square law detectors for each antenna. All visibilities are accumulated for some time in the correlators and are sent to the processor where they are accumulated in a memory. I.e., the accumulation time consists of two parts: the correlator summation number and the software summation number. Usually the correlator/software summation-number ratio is in the range $10^4$-$10^5$, while the accumulation time is in the range 10-100~ms. The 10-antenna correlator IP core together with the embedded processor utilize only about $20\%$ of the EP2S60F1020C4 chip, so that the 100-antenna correlator would not be expensive. 

\subsection{Software and data formats}
Software of the 10-antenna prototype consists of three parts: two applications under $\mu$C/OS-II (the digital receivers and the correlator) and the Linux application for the data store and the observation control. Data are stored both in FITS and in MS2 format. The MS2 format aims to provide compatibility with the CASA package while the FITS files are to be processed with any software. A detailed description of these formats is accessible via ftp\footnote[1]{ftp://badary.iszf.irk.ru/Public/software}.
The Linux and $\mu$C/OS-II applications communicate together using the TCP/IP data transmission protocol. The digital receiver applications are initialized just before start of observations. The Linux application sends to each digital receiver application the RF frequency list, the LO frequency list and the solar ephemerides. The LO frequency list and the accumulation time are sent to the correlator application. During observation the digital receiver application calculates the geometrical delays and phases of the NCO and sends commands to the receiver to correct them. The time interval between the corrections is 7~ms and the fringe frequency of the phase to be corrected is not more than 2-4~Hz. The correlator application acquires data, accumulates them and sends accumulated samples to the Linux application. Also the correlator application lists the LO frequencies by controlling the LO over the SCPI communication protocol.

Both correlator and digital receiver applications consist of multiple tasks. The correlator application tasks are the network support, the GPS time synchronization and the acquisition. The digital receiver application consists of similar three tasks and the tracking task. The tracking task calculates the geometrical delay for the given antenna and prepares the data for the phase correction to be done by the interrupt handler every 7~ms.

The Linux application that controls the observation and stores data is a simple Qt application with network support and graphical interface. Also a Python script is used to plot current data and transmit the plot to the observatory Web server.

\section{Observations}
\begin{figure}
\caption{The flare at NOAA 11164 on 7 March 2011. a) microwave fluxes and spatial sizes of burst source estimated in assuming a symmetrical source. b)~the HXR count rates and microwave brightness temperatures. One can see that each pulse of the microwave flux has a different behavior in consistent with both the HXR counts rate and the microwave source size.} \label{obs_flux_and_temp}
\centerline{\includegraphics[width=\maxfloatwidth]{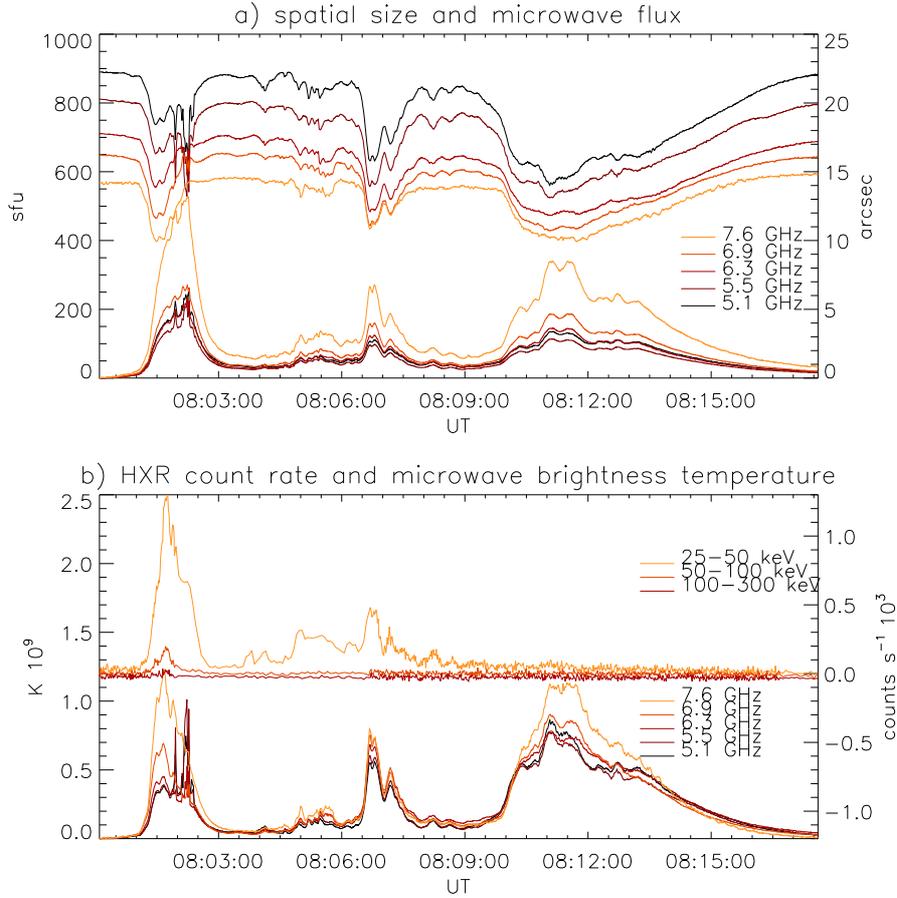}}
\end{figure}
Although the prototype is aimed to find out the answers on technical questions, nevertheless its ability to obtain spatial data simultaneously at a number of frequencies is very important for studies of solar activity. Regular observations with the prototype were started in August 2010. The observations are carried out at frequencies, listed in the configuration file. The temporal resolution is also defined by a record in this file. The temporal resolution is defined as $N_{freqs}\tau_{pol}n$, where $N_{freqs}$ is the frequency list length, $\tau_{pol}$ is the half period of the polarization switching and $n$ is a number of software summations. The typical value of the temporal resolution is in the range 1-2~s. 

A flare of GOES class M1.4 occurred in NOAA 11164 on 7~March 2011. The observations were made at five frequencies (5.1, 5.5, 6.3, 6.9, 7.6~GHz) with temporal resolution of 1.4~s. The flux time profile of the microwave burst is the sequence of pulses at times 08:02, 08:07 and 08:11~UT (Figure~\ref{obs_flux_and_temp}). The maximum of the flux was at 08:02~UT. The event with fine temporal structure~(FTS) occurred between 08:01:55 and 08:02:25~UT. The SSRT has the ability to obtain the displacement of the FTS event with respect to the background burst location or to the nearest quiet source \cite{LesKardSolPh, Meshalkina04}. One can obtain from the SSRT record of this flare that the source of the FTS at 5.7~GHz was displaced to the North-West by about $25^{\prime\prime}$ with respect to the background burst source location. The polarization degree of the FTS was very low.

Under the assumption of an azimuthally symmetrical burst source one can neglect 2D features of its spatial spectrum. In such a case one can use $S(\sqrt{u^2 + v^2})$ instead of the $S(u,v)$, where $S$ is the spatial spectrum and $u$, $v$ are spatial frequencies. The source size can be estimated from the approximation of the 1D spatial spectrum. Both linear or Gaussian approximations are appropriate to estimate the half size of the symmetrical source. The size $\theta$ is defined as $\alpha A^{linear}_1/A^{linear}_0$ or $\beta/(\pi\sqrt{2}A^{Gauss}_2)$ for linear and Gaussian cases respectively, where $A_{k}$ are approximation coefficients. The empirical constants are $\alpha = 0.75$ and $\beta = 1.5$ in assumption of the Gaussian shape of the source. In practice, the linear approximation gives more stable results. This is because the 45 points in the spatial spectrum of the prototype array are grouped into four clusters. And the minimal point number for the successful Gaussian fitting is four points. Therefore the noise influence on the result is stronger for the Gaussian fitting with respect to the linear fitting.

The estimation by using this approximation is valid only for the compact sources. More precisely, if the  longest baseline responses are near zero the estimated source size just follows the spatial resolution of the prototype at the given frequency. During the flare the compact sources appear, and in turn the longest base responses grow and the estimated source size becomes more and more reliable.
Simulation shows that at frequency of 6~GHz and an hour angle of $60^{\circ}$  the linear fitting gives a reliable result for the source size less than $20^{\prime\prime}$, while for the Gaussian fitting the corresponding size is $25^{\prime\prime}$. Also the simulation shows jumps in the Gaussian fitting result. Although the range of reliable sizes is wider for the Gaussian fitting, it is very difficult to obtain a stable result in the presence of noise. In short one can say that the above estimations are valid for the unresolved sources only. If unresolved sources do not appear during a flare this method will be invalid. Obviously, if the source is very small, it is impossible to estimate its size by using this method. The lower limit of source size is about $2-4^{\prime\prime}$, below which the baseline responses become approximately equal.

\begin{figure}
\caption{Correlated phases of 1W-2W, 2W-3W, 127E-128E (left column) and 129S-130S, 130S-131S, 131S-132S (right column) pairs at different frequencies. The diamonds, triangles and squares show the pair phases while solid line shows the averaged phase. The sharp peaks coincide with fine temporal structure events showed in Figure~\ref{obs_flux_and_temp}.} \label{obs_averaged_phase}
\centerline{\includegraphics[width=\maxfloatwidth]{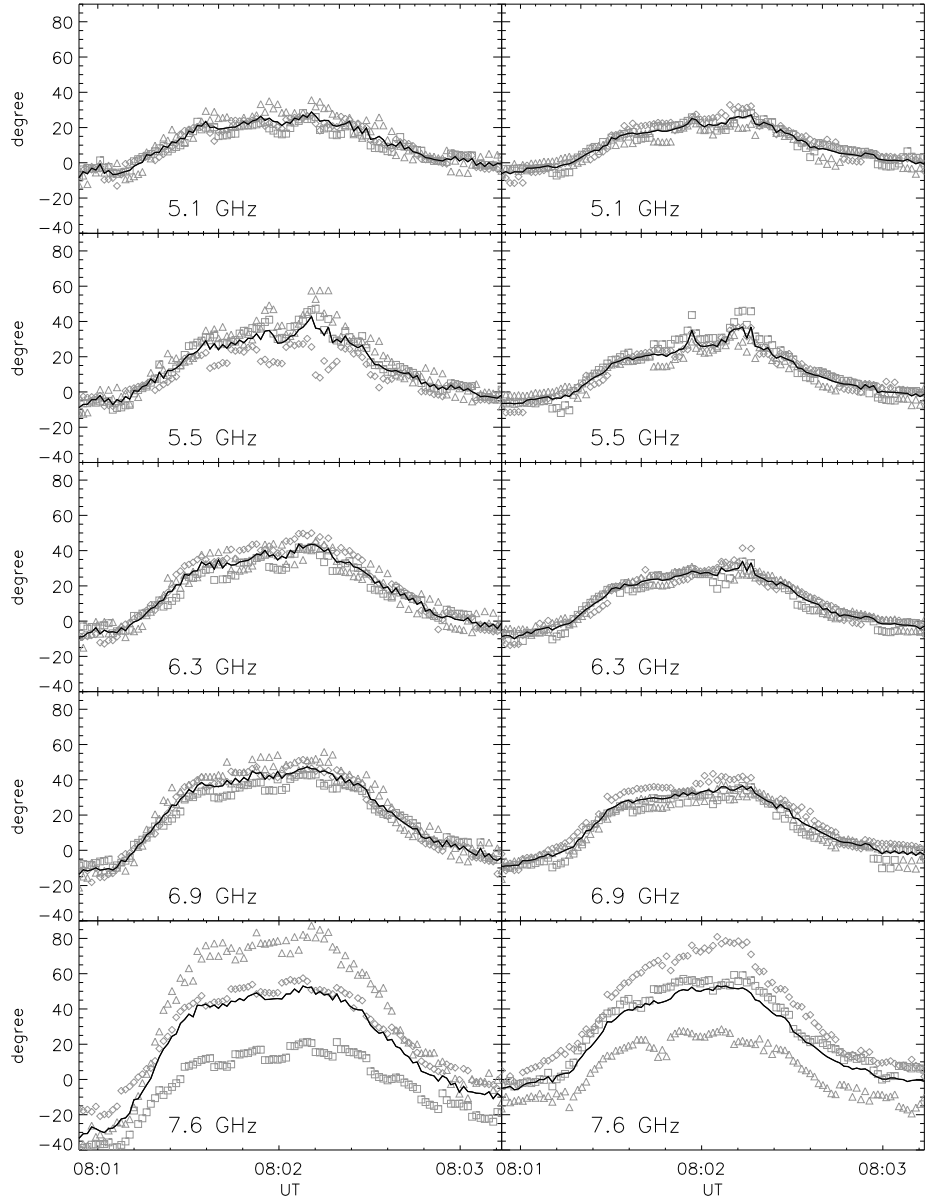}}
\end{figure}

The sizes estimated by using above method are shown on the top in Figure~\ref{obs_flux_and_temp}. The minimum source size was seen during the last pulse in the time profile. One can see that mainly the size decreases while the flux grows. On the other hand, it is distinctly seen that at times between 08:01:30 and 08:02:30~UT the  size grows like the flux. There are three different behaviors of the size during the burst. During the first flux pulse the size first decreases and then grows while the flux is only growing. For the second pulse one can see that the size behavior is anticorrelated with to the flux changes. Finally, during the last flux pulse the size is almost constant especially at high frequencies. It could be explained in the following way. First one can estimate the brightness temperatures during the burst. Figure~\ref{obs_flux_and_temp} shows the brightness temperature obtained from fluxes and sizes by using $T_b(\nu) = 1.39\times10^{10}S(\nu)/\left(\nu\theta(\nu)\right)^2$, where $\nu$ is the frequency in GHz, $\theta(\nu)$ is the size in arcseconds and $S(\nu)$ is the observed flux in sfu. There is some discrepancy between the brightness temperatures measured with the SSRT and estimated with the prototype. The SSRT data show that the brightness temperature near 08:02~UT was $2.2\times10^7$~K (5.7~GHz) while the estimated brightness temperature at 5.5~GHz was of $3\times10^8$~K (Figure~\ref{obs_flux_and_temp}). But the SSRT beam and estimated source sizes should be taken in account. The corrected brightness temperature at 5.7~GHz is about $10^8$~K. 

It is generally believed that the microwave and hard X-ray emission are produced by the same population of particles \cite{HoltRamaty, BastianBenzGary}. So, it would be useful to combine the microwave data with the hard X-ray~(HXR) data. We used data from the \textit{Gamma-Ray Burst Monitor}~(GBM) onboard the \textit{Fermi} satellite\footnote[1]{ http://hesperia.gsfc.nasa.gov/fermi\_solar}. As can see from the Figure~\ref{obs_flux_and_temp} the HXR emission is observed during only the first two pulses of the burst. Moreover, the high energy accelerated particles (50-100~keV) are recorded during the first pulse only. Therefore, one can say that the microwave burst began with appearance of the high energy particles at around 08:01 UT. The brightness temperature followed the HXR emission, but the size increased quickly with respect to the temperature decreasing. In turn, the microwave flux increased during the first pulse of the burst. The relatively rapid fluctuations of the size during the second pulse coincide with the HXR emission ones. Finally, only weak variations of the source size are observed during the third pulse when the HXR emission has decreased. 

\begin{figure}
\caption{Displacements of fine temporal structure event sources at different frequencies with respect to the location of the background burst source at 5.7 GHz. Contours show the image of the background burst source at 5.7 GHz (SSRT). The square indicates the SSRT detected displacements. Colored triangles are displacements obtained with the 10-antenna prototype at different frequencies.} \label{obs_displacements_ssrt}
\centerline{\includegraphics[width=\maxfloatwidth]{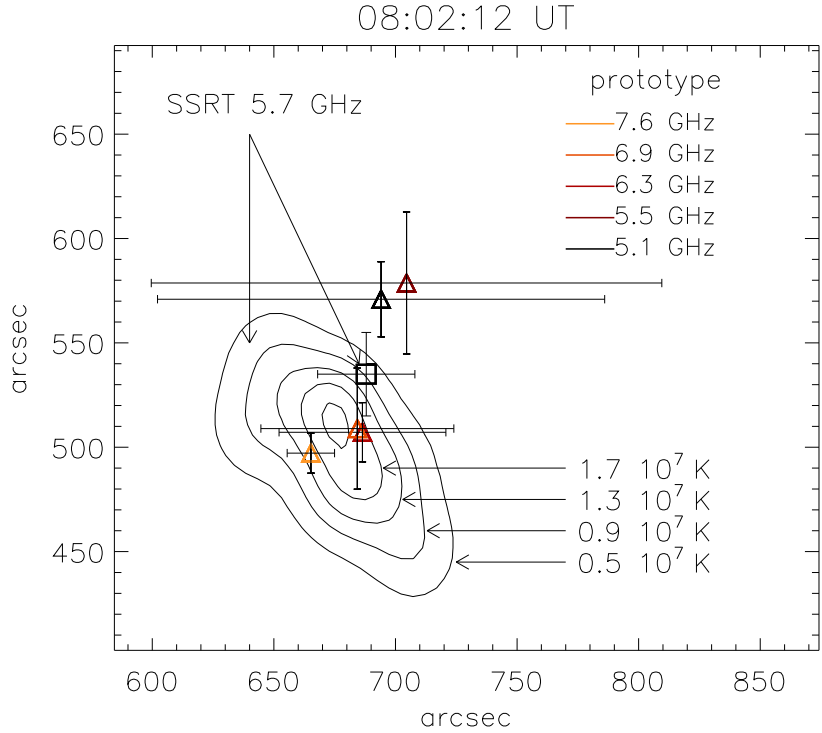}}
\end{figure}

The FTS event near 08:02 UT has very different brightness temperature and spatial size (Figure~\ref{obs_flux_and_temp}) with respect to the background burst. The FTS temperature and size peaks near 5.5~GHz while the background peaks at 7.6~GHz. As mentioned above, the FTS source is spatially shifted relative to the background burst brightness center at the SSRT frequency of 5.7~GHz. On the other hand, one can estimate the source displacement during the burst by using the 10-antenna prototype data. The mean phases of the shortest baselines of the East-West and North-South prototype subarrays are used for this estimation. There are four such baselines in the East-West array and three baselines in the North-South array. The mean phases during the burst are shown in Figure~\ref{obs_averaged_phase}. One can neglect the slow variations of the phase to find out the FTS displacements. In such a case the displacement could be inferred from phases by simple relations $\Delta\xi = \Delta\phi_{kl} / (2\pi |\vec{b}_{\lambda}^{kl}(t)|)$ and $\Delta\eta = \Delta\phi_{mn} / (2\pi |\vec{b}_{\lambda}^{mn}(t)|)$, where $\Delta\phi$ phases and $\vec{b}_{\lambda}(t)$ baselines of the $(k,l), (m,n)$ antenna pairs respectively. The picture plane displacement can be inferred from $(\Delta\xi$,$\Delta\eta)$ taking into account that $(\Delta\xi$,$\Delta\eta)$ are covariant coordinates in an oblique coordinate system \cite{Lesovoy1999}:

\[ 
\left( \begin{array}{c}
\Delta x \\ 
\Delta y
\end{array}\right) =  \frac{1}{\sin^2{\gamma}} \left( \begin{array}{cc}
	\cos{\gamma_{kl}} & \cos{\gamma_{mn}} \\
	\sin{\gamma_{kl}} & \sin{\gamma_{mn}} 
\end{array} \right)\left( \begin{array}{cc}
	1 & -\cos{\gamma} \\
	-\cos{\gamma} & 1 
\end{array} \right)\left( \begin{array}{c}
\Delta \xi \\ 
\Delta \eta
\end{array}\right)
\]
where $\gamma_{kl}$, $\gamma_{mn}$ are angles between normal to the baselines and the normal to the solar trajectory and
\[ 
\gamma=\left\{
\begin{array}{cc}
\gamma_{kl} - \gamma_{mn},&	 h \leq 0 \\ 
\pi - \left(\gamma_{kl} - \gamma_{mn}\right),&  h > 0
\end{array}
\right.
\]
an angle of the oblique coordinate system, $h$ is the hour angle.

Figure~\ref{obs_averaged_phase} shows the visibility phases for the first pulse of the burst. The sharp peaks of the phase at three lower frequencies coincide with the FTS events showed in Figure~\ref{obs_flux_and_temp}. One can suggest that these peaks are due to the displacement of the FTS position with respect to the background burst. But the peak values are too high to be explained by the displacement only. The flux variation and the displacement can affect the visibility phase simultaneously. To reduce influence of the the background flux variation it was subtracted from the spatial spectrum. Figure~\ref{obs_displacements_ssrt} shows the image of the background burst source at frequency 5.7~GHz obtained by the SSRT and displacements of the FTS at different frequencies obtained from the visibility phases of the subtracted spectrum. Two baselines were used for each measurement of the displacement. After that all results were averaged. The square marks the locations of the FTS source detected by the SSRT. The locations of the FTS sources measured with the prototype are shown by colored triangles. The scattering of displacements is  higher than that obtained by the SSRT. The prototype displacements at frequencies 5.1 and 5.5 GHz are two times greater than the SSRT displacement at 5.7~GHz. This is likely due to a residual influence of the background burst. Nevertheless, the consistency between these displacements is distinctly seen. Also one can see that the displacements depend on frequency.

\section{Conclusions}
The description of the 10-antenna prototype of the multifrequency Siberian radioheliograph is presented. All parts of the prototype are detailed including  antenna with broadband front-end, downconvertor, analog receiver, digital receiver and correlator. The control and data acquisition software are also described. The main goal of the 10-antenna prototype project was to answer the technical questions on the way to developing the multifrequency Siberian radioheliograph. All the major questions were answered and the start of the test observation of the 96-antenna multifrequency Siberian radioheliograph is scheduled for autumn of 2012.

The analysis of the solar burst data obtained by the 10-antenna prototype shows the impressive capabilities of simultaneous observations with spatial, spectral and temporal resolutions. In other words, one can say that the analysis eliminates the problems with studying solar flares without the spatial resolution in microwaves. One can see from the 7~March 2011 flare observation that joint analysis the HXR and microwave spectral data without spatial data gives ambiguous result. On the other hand, simple adding to the analysis of an information on microwave source sizes changes the whole picture. The sources of fine temporal structure events often are displaced with respect to the background burst source location. Also spectra and sizes of such event are different with respect to the background burst ones. The ability of the 10-antenna prototype (and the future heliograph) to study such events seems to be very important because generally believed that such events are manifestation of the primary energy releases.

\acknowledgements This research was supported by  the Russian Foundation of Basic Research  grants 12-02-91161-$ГФЕН$a, and 12-02-00173-a and grants of Federal Agency for Science and Innovation (State Contracts 16.518.11.7065 and 02.740.11.0576). We thank Larisa K. Kashapova for help with preparing the HXR data. We would like to thank the anonymous referee for the helpful comments and suggestions.

\end{article}
\end{document}